\documentstyle{JFM}
\input{psfig}

\newcommand{\be}{\begin{equation}}
\newcommand{\ee}{\end{equation}}
\newcommand{\bea}{\begin{eqnarray}}
\newcommand{\eea}{\end{eqnarray}}

\newcommand{\vx}{{\bf x}}

\newcommand{\hkj}{{\hat{k}_j}}
\newcommand{\hkl}{{\hat{k}_l}}

\newcommand{\vk}{{\bf k}}

\newcommand{\sF}{{\cal F}}

\newcommand{\sD}{{\hat{{\cal D}}}}

\title[Pattern formation in weakly damped  ...]
{Pattern formation in weakly damped parametric
surface waves driven by two frequency components}
\author[WENBIN ZHANG AND JORGE VI\~NALS]{WENBIN ZHANG$^1$ AND JORGE
VI\~NALS$^2$}
\affiliation{$^1$Department of Chemical Engineering,
             Massachusetts Institute of Technology, Cambridge,
             Massachusetts 02139, USA\\
             $^2$Supercomputer Computations Research Institute,
             Florida State University, Tallahassee, Florida 32306-4052, USA,
             and Department of Chemical Engineering,
             FAMU-FSU College of Engineering, Tallahassee, Florida 31310-6046,
USA}
\pubyear{1996}
\volume{000}
\pagerange{000--000}
\date{\today}
\begin{document}
\maketitle

\begin{abstract}
A quasi-potential approximation to the Navier-Stokes equation for low
viscosity fluids is developed to study
pattern formation in parametric surface waves 
driven by a force that has two frequency components. A bicritical line 
separating regions of instability to either one of the driving 
frequencies is explicitly obtained, and compared with experiments 
involving a frequency ratio of 1/2. The procedure for deriving
standing wave amplitude equations valid near onset is
outlined for an arbitrary frequency ratio following a multiscale asymptotic
expansion of the quasi-potential equations. Explicit results
are presented for subharmonic response to a driving
force of frequency ratio 1/2, and used to study pattern selection.   
Even though quadratic terms are prohibited in this case, hexagonal
or triangular patterns are found to be stable in a relatively large
parameter region, a fact that is in qualitative agreement with 
experimental results.
\end{abstract}

\section{Introduction}
\label{sec:introduction}

When a fluid layer is periodically oscillated in the direction normal 
to the free surface at rest, parametric surface waves (or Faraday
waves) appear above a certain critical value of the vibration
amplitude (Miles \& Henderson 1990; Cross \& Hohenberg 1993). 
We present in this paper 
an extension of a weakly nonlinear model previously introduced 
by Zhang \& Vi\~nals (1996a) and (1996b), 
which is valid in the limit of low fluid 
viscosity, large aspect ratio and large depth, to study pattern 
selection near onset of Faraday waves when the driving force has 
two independent frequency components. 

The model developed by Zhang \& Vi\~nals (1996b) was based on 
a quasi-potential approximation to the equations governing fluid 
motion.  In it, the flow is considered to be potential in the bulk, 
subject to effective boundary conditions at the moving free
surface that incorporate the effect of the rotational component of the 
flow within a thin boundary layer near the free surface.  We further 
assumed without rigorous justification that for low viscosity fluids,
only linear viscous terms need to be retained in the resulting
equations (the so-called \lq\lq linear damping quasi-potential
equations", or LDQPE's).  A multi-scale analysis of the resulting 
LDQPE's led to the prediction of standing
wave patterns of square symmetry near onset for capillary waves, 
in agreement with experiments.  For mixed capillary-gravity waves, 
patterns of hexagonal symmetry or quasi-periodic patterns were 
predicted depending on the value of the damping coefficient.

Although some of our predictions have been confirmed experimentally
(Kudrolli \& Gollub 1996), we address in this paper pattern selection in 
systems driven by periodic forces comprising two frequency components 
for two reasons. First, recent detailed experiments involving two 
frequencies provide a good opportunity for additional tests
of our theory. Second, additional control 
parameters appear relative to the single frequency case, namely, the 
frequency ratio, relative amplitude and phase difference.  
Therefore, one may anticipate richer dynamics and more interesting 
steady states as compared to the single frequency case.  

Consider a fluid layer perpendicular to the $z$ axis, driven by a 
force of the form
\begin{equation}
\label{eq:force}
g_{z}(t) = -g_0 - g_{z}[r\sin 2m\omega_0 t + (1-r) \sin (2n\omega_0 t + \phi)],
\end{equation}
where $r$ ($0 \le r \le 1$) is the relative amplitude of the two
frequency components, $\phi$ is their relative phase difference, 
$g_{0}$ is the constant background gravitational field pointing 
to the negative $z$ direction, and $g_{z}$ is the amplitude of the 
oscillatory component.  When $r$ is sufficiently small, the forcing 
component $\sin (2n\omega_0 t+\phi)$ is expected to dominate, the 
other component ($\sin 2m\omega_0 t$) being a small perturbation. 
Linear stability of the flat surface is thus approximately determined by 
subharmonic instability to the forcing component $\sin (2n\omega_0 t+\phi)$
(note that this is not necessarily a subharmonic response to the
entire driving force), with a critical wavenumber 
$k_{0n}$ given by $g_0k_{0n} + \Gamma k_{0n}^3/\rho = n^2\omega_0^2$, 
where $\Gamma$ is the surface tension of the surface and $\rho$ is the 
density of the fluid.  On the other hand, when $r$ is sufficiently 
close to 1, the forcing component 
$\sin 2m\omega_0 t$ dominates,  and linear stability is approximately
determined by subharmonic instability at a frequency $m\omega_0$ (again
it is not necessary that it be a subharmonic response to the entire driving 
force), with a critical wavenumber $k_{0m}$ given by $g_0k_{0m} + 
\Gamma k_{0m}^3/\rho = m^2\omega_0^2$.  When $r$ is varied from 0 to 1, 
one could imagine that the above two instabilities can have the same
threshold value 
of the driving amplitude $g_{z}$ for some value of $r$, which will be 
referred to as the {\em bicritical} value $r_b$. Furthermore, given 
a fixed frequency ratio 
$m/n$, the bicritical value $r_b$ can be a function of the phase difference 
$\phi$.  In the $r-\phi$ plane, the bicritical values $r_b(\phi)$ form a line
(the {\em bicritical line}) which separates the two regions with different 
characteristic temporal dependencies and spatial scales. 

The first two-frequency Faraday experiments were reported by Edwards 
and Fauve (Edwards \& Fauve 1992, 1993 and 1994), albeit in
fluids of large viscosity.  In that case, and
with a purely sinusoidal parametric force, roll patterns (or lines)
similar to those observed in Rayleigh-B\'enard convection are observed 
near the primary instability of a flat surface.  In the case of two 
frequencies,
Edwards and Fauve studied several ratios $m/n$, including 3/5, 4/5,
4/7, 6/7, and 8/9, with most of their data for the case  $m/n=4/5$.  
In this latter case, parametric surface waves were found to
respond synchronously (harmonically) with the force when the even frequency 
forcing component 
$\sin (2\times 4\omega_0t)$ dominates, and subharmonically when the
odd frequency forcing component $\sin (2\times 5\omega_0t + \phi)$
dominates.  The bicritical line obtained by Edwards and Fauve is almost
independent of the phase difference $\phi$, and for $m/n=4/5$, $r_b(\phi) 
\approx 0.32$.  

They also found many interesting 
standing wave patterns near the primary instability of a planar surface: 
lines, squares, hexagons, and twelve-fold quasi-crystalline patterns.  
For the cases of $m/n = \mbox{even/odd}$ (4/5, 4/7, 6/7, and 8/9), 
hexagonal patterns were observed for a wide range of values of $r$,
in the region of harmonic response in the $r-\phi$ plane (when the even 
frequency forcing component dominates).
On the same side of the bicritical line where hexagonal patterns are found, 
a stable twelve-fold quasi-crystalline pattern was observed in a small  
region very close to the bicritical line.  For $m/n = 4/5$, this small region
is around $r \approx 0.32$ and $\phi \approx 7.5^{\circ}$.
\footnote{In the form
of the driving force used by Edwards and Fauve, 
$-g_0 + a_0[\cos\chi\cos(4\omega t) + \sin\chi\cos(5\omega t + \phi')]$, 
the twelve-fold quasi-crystalline pattern was
observed around $\chi \approx 65^{\circ}$ and $\phi' \approx 75^{\circ}$.
Also, for later reference, we note that \cite{re:muller93} used the
definition
$g(t) = -g_0 + g_{z}[r\cos 2\omega_0 t + (1-r) \cos (4\omega_0 t +\varphi)]$,
which can be written in the form of 
Eq.~(\ref{eq:force}) by the transformation of $t \rightarrow t+\pi/(4\omega_0)$
and $\varphi = \phi - \pi/2$.}

Two-frequency driven Faraday experiments using less viscous fluids have 
been reported more recently by M\"uller (1993).  In the case of 
single frequency forcing, M\"uller observed standing wave patterns 
of square symmetry near onset, in agreement with previous experiments 
(Lang 1992; Ezerskii, Rabinovich, Reutov \& Starobinets 1986; 
Tufillaro \& Gollub 1989; Ciliberto, Douady \& Fauve 1991;
Christiansen, Alstr{\o~}m \& Levinsen 1992; Bosch \& van de Water 1993;
Edwards \& Fauve 1993)
and our previous theoretical work (Zhang \& Vi\~nals 1996b) for weak viscous
dissipation.  He also studied 
pattern formation near onset in a system driven by a two-frequency 
force, with frequency ratio $m/n = 1/2$, and
found that the $r-\varphi$ plane is divided into two regions: in
the region of larger $r$, parametric surface waves are found to respond 
subharmonically to the entire driving force, whereas in the region of smaller
$r$ surface waves respond harmonically. The bicritical line has an interesting 
dependence on the phase difference $\varphi$, which was not observed by 
Edwards and Fauve (1993) and (1994) for frequency ratios other 
than $1/2$ in highly viscous fluids.

In addition to square patterns, M\"uller found hexagonal patterns that 
respond harmonically to the driving force, and 
hexagonal or triangular patterns that respond subharmonically to the 
driving force.  Interestingly, a spatially disordered region, that
responds subharmonically to the driving force with an additional slow time 
dependence, was also found near onset.  

We study in this paper how much of this seemingly complicated 
stability diagram found by M\"uller can be understood, at least 
qualitatively, in terms of perturbative analysis of the 
quasi-potential equations introduced by Zhang \& Vi\~nals (1996a) and
(1996b).  
In M\"uller's experiments, the damping parameter ($\gamma = 2 \nu k^{2}$,
where $\nu$ is the kinematic viscosity of the fluid, and $k$ the
wavenumber of a surface Fourier mode) 
$\gamma_s=0.17$ in the region of subharmonic response, and 
$\gamma_h=0.33$ in the region of harmonic response.  
Since the quasi-potential approximation is only valid in the limit
of small viscous dissipation (small $\gamma$), we anticipate qualitative 
agreement with experiments, while smaller damping parameters are needed
to quantitatively test our theoretical predictions.  

The rest of this paper is organized as follows.  In section 
\ref{sec:bicritical}, we present analytical results for the bicritical
lines, and make comparison with M\"uller's experimental results.
Section \ref{sec:SWAE_two_frequency} contains the derivation of 
standing wave amplitude equations (SWAE's), and results of pattern 
selection based on the SWAE's.  Further discussion and conclusion are
presented in the section \ref{sec:discussion}.
  
\section{Bicritical line}
\label{sec:bicritical}

In this section, we consider the linear stability of two-frequency forced 
Faraday waves, and obtain analytical results for the bicritical line 
$r_b(\phi)$.
We restrict our attention to cases in which $m$ and $n$ are two small 
positive integers that are relatively prime (e.g, $m/n = 1/2, 1/3, 2/3$). 
Without loss of generality, we assume $m < n$.  The phase difference $\phi$
can be chosen within $-\pi/m \le \phi < \pi/m$ because $g(t)$ is 
invariant with respect to the transformation: $\phi \rightarrow 
\phi + 2\pi/m$ and $t \rightarrow 
t + j\pi/m\omega_0$ for any integer $j$ that makes $(nj+1)/m$ an integer.
Linear stability of the free surface of a weakly viscous fluid of 
infinite depth under a two-frequency parametric 
force is determined by the damped Mathieu equation (Landau 
\& Lifshitz (1976)),
\be
\label{eq:h2_12}
\partial_{tt}\hat{h}_k + 4\nu k^2\partial_t \hat{h}_k
+\Biggl[g_0 k + \frac{\Gamma k^3}{\rho} + g_{z}k\biggl(r\sin 2m\omega_0t
+ (1-r)\sin(2n\omega_0 t + \phi)\biggr) \Biggr]\hat{h}_k = 0.
\ee

\subsection{Subharmonic versus synchronous responses}
\label{sec:subharmonic_sinchronous}

The two-frequency parametric force in Eq.~(\ref{eq:h2_12}) has an 
angular frequency $2\omega_0$, or a period $\pi/\omega_0$ (recall that $m$ 
and $n$ are assumed to be relatively prime).  Whether a region is 
subharmonic or synchronous (harmonic) with respect to the total
driving force depends on whether $m/n$ is
odd/odd, even/odd, or odd/even.  This overall response to the total force
is important since it can be easily determined experimentally by using
stroboscopic methods.
According to Floquet's theorem, the general form of the solutions of 
Eq.~(\ref{eq:h2_12}) is a periodic function multiplied by an exponential 
function of time. At the stability boundaries, 
Eq.~(\ref{eq:h2_12}) has periodic solutions,
$ \hat{h}_k(t + \pi/\omega_0) = p \hat{h}_k(t),$
where the Floquet multiplier $p = \pm 1$.  For $p = -1$, $\hat{h}_k(t)$
has a period of $2\pi/\omega_0$, and thus the response is subharmonic,
whereas for $p = 1$, $\hat{h}_k(t)$ has a period of 
$\pi/\omega_0$, and hence the response is harmonic.
Therefore the solution of Eq.~(\ref{eq:h2_12})
at the boundaries for subharmonic instability ($p = -1$) can be
written as a Fourier series in odd frequencies $ (2j-1) \omega_{0},
j=1,2, \ldots , \infty$, whereas at the boundaries for harmonic
instability ($p = 1$) the frequencies involved in the series are
$2 j \omega_{0}, j=1, 2, \ldots , \infty$. As a consequence, when the
odd (even) frequency dominates, the response is subharmonic (harmonic).

We now turn to a detailed calculation of the bicritical line in the
$(r, \phi)$ plane, starting from the linearized quasi-potential equation
(Eq. (\ref{eq:h2_12})) (Zhang \& Vinals 1996a and 1996b).
Since the characteristic time and length scales are different on 
the two sides of the bicritical line for two-frequency driven Faraday waves, 
we shall use dimensional variables, and group them into 
dimensionless quantities when necessary.  In order to keep our notation 
simple, we define 
$\delta = g_0 k +\Gamma k^3/\rho$, and $f=g_{z}k/4$.  
Equation (\ref{eq:h2_12}) can now be written as
\be
\label{eq:h2_12_}
\partial_{tt} \hat{h}_k + 2\gamma\partial_t \hat{h}_k 
+\Biggl[\delta + 4f\biggl(r\sin 2m\omega_0t + (1-r)\sin(2n\omega_0 t +
\phi)\biggr)\Biggr] \hat{h}_k = 0.
\ee

\subsection{Multiple scale expansion}
\label{sec:multiple_scale}

We consider small 
values of the driving amplitude $f$ and of the damping coefficient $\gamma$,
and introduce a small parameter $\eta$ such that $ \gamma = \eta
\gamma_0$ and $f = \eta f_0$.  
For simplicity, we will consider the wavenumber $k_0$ exactly at
subharmonic resonance to either of the forcing components, and thus no 
expansion for the wavenumber is needed, define $\delta_0 = g_0k_0 + \Gamma
k_0^3/\rho$, and
seek a solution $\hat{h}_k$ in a power series in $\eta$ as,
\be
\hat{h}_k = \hat{h}_k^{(0)}(t, T_1, T_2) + \eta \hat{h}_k^{(1)}(t, T_1, T_2) 
+ \eta^2  \hat{h}_k^{(2)}(t, T_1, T_2) + \cdots,
\ee
where $T_1 = \eta t$ and $T_2 = \eta^2 t$.  We have introduced a second
slow time scale $T_2$ in the expansion for $\hat{h}_k$, in addition to the slow
time scale $T_1$. The second slow time scale is necessary since we are 
going to perform our 
perturbation expansion up to ${\cal O}(\eta^2)$, which is one order higher 
in $\eta$ than the perturbation expansion for the case of a single 
sinusoidal driving force in \cite{re:zhang96b}. 

    Equation (\ref{eq:h2_12_}) contains a damping term proportional 
to $\gamma$, 
while terms of ${\cal O}(\gamma^{3/2})$ and ${\cal O}(\gamma^{2})$ (or 
${\cal O}(\eta^2)$) have been neglected 
in the quasi-potential approximation (Zhang \& Vi\~nals 1996b). 
It seems to be inconsistent to consider terms of ${\cal O}(\eta^2)$ in
the solutions to Eq.~(\ref{eq:h2_12_}).  The basic reason to
consider solutions up to ${\cal O}(\eta^2)$ here is that, as we shall 
see below, these terms can have relatively large coefficients due to the 
special nature of the expansion. When the dominant driving frequency is
$2m\omega_0$, the small parameter in this perturbation expansion is
the dimensionless driving amplitude $rf/(m^2\omega_0^2) \ll 1$.
The driving force in this case can be written as
\be
\frac{4rf}{m^2\omega_0^2}\left(\sin 2m\omega_0t 
               + \frac{1-r}{r} \sin(2n\omega_0t + \phi)\right).
\ee
Since the perturbative expansion is in be in powers of
$rf/(m^2\omega_0^2)$, as well as in $(1-r)/r$, it is necessary to assume that
$(1-r)/r \sim {\cal O}(1)$ or smaller. However, $(1-r)/r$ can
be quite large on the bicritical line if $r= r_b(\phi)$ is small
for some values of $\phi$.
In this case, higher order terms in $\eta$ can be important if they have factors
of $(1-r)/r$ in their coefficients. 
In M\"uller's experiments, for example, the smallest value of $r_b(\phi)$ 
is about 0.2, and thus $(1-r)/r \sim 4$. Similar arguments can be made
when $2n\omega_0$ is the dominating driving frequency.
In the case of M\"uller's experiments,
the largest value of $r/(1-r)$ along the bicritical line is less than 0.5.  
A consistent calculation would require keeping terms of 
${\cal O}(\gamma^{3/2})$ or higher at the linear level of approximation
in the surface variables (Eq. (\ref{eq:h2_12_})), leading to a much more
involved nonlinear analysis. However, the agreement that we find with
M\"uller's bicritical line is quite reasonable.

On substituting the expansion for $\hat{h}_k$ into Eq.~(\ref{eq:h2_12_}), 
we have at ${\cal O}(\eta^0)$,
\be
\partial_{tt} \hat{h}_k^{(0)} + \delta_0 \hat{h}_k^{(0)} = 0.
\ee
At ${\cal O}(\eta)$, we have,
\be
\partial_{tt} \hat{h}_k^{(1)} + \delta_0 \hat{h}_k^{(1)} 
= -2 \frac{\partial^2 \hat{h}_k^{(0)}}{\partial T_1 \partial t}
  -2\gamma_0 \frac{\partial \hat{h}_k^{(0)}}{\partial t}
  -4f_0\biggl(r\sin 2m\omega_0t + (1-r)\sin(2n\omega_0 t +\phi)\biggr)
   \hat{h}_k^{(0)},
\ee
and at ${\cal O}(\eta^2)$, we have,
\bea
\partial_{tt} \hat{h}_k^{(2)} + \delta_0 \hat{h}_k^{(2)}
= -2 \frac{\partial^2 \hat{h}_k^{(0)}}{\partial T_2 \partial t}
  -  \frac{\partial^2 \hat{h}_k^{(0)}}{\partial T_1^2}
  -2\gamma_0 \frac{\partial \hat{h}_k^{(0)}}{\partial T_1}
  -2\frac{\partial^2 \hat{h}_k^{(1)}}{\partial T_1 \partial t}
  -2\gamma_0 \frac{\partial \hat{h}_k^{(1)}}{\partial t}
\nonumber \\
  -4f_0\biggl(r\sin 2m\omega_0t + (1-r)\sin(2n\omega_0 t +\phi)\biggr)
   \hat{h}_k^{(1)}.
\eea
We now consider separately the two cases of $\delta_0 = m^2\omega_0^2$
and $\delta_0 = n^2\omega_0^2$.

\subsubsection{{$\delta_0 = m^2\omega^2_0$}}
At ${\cal O}(\eta^0)$,
we have the solution $\hat{h}_k^{(0)} = A(T_1, T_2) \cos m\omega_0t 
+ B(T_1, T_2) \sin m\omega_0t$.  At ${\cal O}(\eta)$, we have the following
solvability condition in order to avoid secular terms,
\bea
\frac{\partial A}{\partial T_1} = \left(\frac{rf_0}{m\omega_0}-\gamma_0\right)A,
\\
\frac{\partial B}{\partial T_1} = -\left(\frac{rf_0}{m\omega_0}+\gamma_0\right)B.
\eea
The solution at this order reads,
\bea
\hat{h}_k^{(1)} = -\frac{rf_0}{4m^2\omega_0^2}
     \left(B\cos 3m\omega_0t - A\sin 3m\omega_0t\right) 
\nonumber \\
    +\frac{(1-r)f_0}{2n(n-m)\omega_0^2}
     \biggl[B\cos\biggl((2n-m)\omega_0t+\phi\biggr) 
          + A\sin\biggl((2n-m)\omega_0t+\phi\biggr)\biggr]
\nonumber \\
    -\frac{(1-r)f_0}{2n(m+n)\omega_0^2}
     \biggl(B\cos\biggl((2n+m)\omega_0t+\phi\biggr) 
          - A\sin\biggl((2n+m)\omega_0t+\phi\biggr)\biggr].
\eea

Whether higher order contributions are important or negligible is not known
at this point.  Since we have found nontrivial equations for the amplitude
$A$ and $B$, and these equations give us a threshold value of the driving 
amplitude $f_0 = m\omega_0\gamma_0/r$ for the Faraday instability, one 
would guess that it is not necessary to consider higher order contributions.  
However, as we show below, contributions at ${\cal O}(\eta^2)$ are 
important to determine the bicritical line $r_b(\phi)$.  Interestingly, and for 
a quite obvious reason to be also discussed below, contributions at 
${\cal O}(\eta^2)$ for the case of $m/n = 1/2$ are qualitatively different from 
the cases of any other frequency ratios.

At ${\cal O}(\eta^2)$, we have,
\bea
\!\!\!&\!\!\!&\!\!\!
\partial_{tt} \hat{h}_k^{(2)} + m^2\omega_0^2 \hat{h}_k^{(2)} =
\nonumber \\
\!\!&\!\!&\!\!
\left[2m\omega_0\frac{\partial A}{\partial T_2} 
       - \frac{\partial^2 B}{\partial T_1^2}
       -2\gamma_0 \frac{\partial B}{\partial T_1}
       -\frac{f_0^2}{\omega_0^2}
        \left(\frac{r^2}{2m^2} + \frac{2(1-r)^2}{n^2-m^2}\right)B
 \right] \sin m\omega_0t
\nonumber \\
\!\!&\!\!&\!\!
-\left[2m\omega_0\frac{\partial B}{\partial T_2} 
       + \frac{\partial^2 A}{\partial T_1^2}
       +2\gamma_0 \frac{\partial A}{\partial T_1}
       +\frac{f_0^2}{\omega_0^2}
        \left(\frac{r^2}{2m^2} + \frac{2(1-r)^2}{n^2-m^2}\right)A 
 \right] \cos m\omega_0t
\nonumber \\
\!\!&\!\!&\!\!
-\frac{2m^2-mn+n^2}{2nm^2(n-m)}\frac{r(1-r)f_0^2}{\omega_0^2}
 \biggl(A\cos[(2n-3m)\omega_0t\!+\!\phi]-B\sin[(2n-3m)\omega_0t\!+\!\phi]\biggr)
\nonumber \\
\!\!&\!\!&\!\!
+\frac{2m^2+mn+n^2}{2nm^2(n+m)}\frac{r(1-r)f_0^2}{\omega_0^2}
 \biggl(A\cos[(2n+3m)\omega_0t\!+\!\phi]+B\sin[(2n+3m)\omega_0t\!+\!\phi]\biggr)
\nonumber \\
\!\!&\!\!&\!\!
-\frac{2(n^2+mn-m^2)}{mn(n^2-m^2)}\frac{r(1-r)f_0^2}{\omega_0^2}
 \biggl(A\cos[(2n-m)\omega_0t\!+\!\phi]+B\sin[(2n-m)\omega_0t\!+\!\phi]\biggr)
\nonumber \\
\!\!&\!\!&\!\!
+\frac{2(n^2-mn-m^2)}{mn(n^2-m^2)}\frac{r(1-r)f_0^2}{\omega_0^2}
 \biggl(A\cos[(2n+m)\omega_0t\!+\!\phi]+B\sin[(2n+m)\omega_0t\!+\!\phi]\biggr)
\nonumber \\
\!\!&\!\!&\!\!
+\frac{(1-r)^2f_0^2}{n(n-m)\omega_0^2}
 \biggl(A\cos[(4n-m)\omega_0t+2\phi]
      - B\sin[(4n-m)\omega_0t+2\phi]\biggr)
\nonumber \\
\!\!&\!\!&\!\!
+\frac{(1-r)^2f_0^2}{n(m+n)\omega_0^2}
 \biggl(A\cos[(4n+m)\omega_0t+2\phi]
      + B\sin[(4n+m)\omega_0t+2\phi]\biggr)
\nonumber \\
\label{eq:order2_12m}
\!\!&\!\!&\!\!
-\frac{3r^2f_0^2}{2m^2\omega_0^2}
 \biggl(A[\cos 3m\omega_0t - \cos 5m\omega_0t] 
      - B[\sin 3m\omega_0t + \sin 5m\omega_0t]\biggr).
\eea
Terms on the RHS of the above equation will contribute to the solvability
condition only if their frequencies are $m\omega_0$.  Thus, besides
the first two
terms, only the third term on the RHS of Eq.~(\ref{eq:order2_12m}) 
contributes to the solvability condition when $2n-3m = m$, i.e., the 
frequency ratio $m/n=1/2$ (note $n>m$ by assumption).  Therefore the  
solvability conditions at ${\cal O}(\eta^2)$ for frequency ratio $m/n=1/2$
is different from that for all other frequency ratios.

For $m/n=1/2$, the solvability conditions read,
\be
\frac{\partial A}{\partial T_2} =
\frac{1}{2\omega_0} \left[\frac{f_0^2}{\omega_0^2}
\left(\frac{3r^2}{2}+\frac{2(1-r)^2}{3}+r(1-r)\cos\phi\right) 
- \gamma_0^2\right]B - \frac{r(1-r)f_0^2}{2\omega_0^3}A\sin\phi,
\ee
\be
\frac{\partial B}{\partial T_2} =
-\frac{1}{2\omega_0} \left[\frac{f_0^2}{\omega_0^2}
\left(\frac{3r^2}{2}+\frac{2(1-r)^2}{3}+r(1-r)\cos\phi\right) 
- \gamma_0^2\right]A + \frac{r(1-r)f_0^2}{2\omega_0^3}B\sin\phi.
\ee
For a frequency ratio $m/n \ne 1/2$, the solvability conditions are, 
\bea
\frac{\partial A}{\partial T_2} = \frac{1}{2m\omega_0} 
\left[\frac{f_0^2}{\omega_0^2}
      \left(\frac{3r^2}{2m^2} + \frac{2(1-r)^2}{n^2-m^2}\right)
      - \gamma_0^2\right]B,
\\
\frac{\partial B}{\partial T_2} = -\frac{1}{2m\omega_0} 
\left[\frac{f_0^2}{\omega_0^2}
      \left(\frac{3r^2}{2m^2} + \frac{2(1-r)^2}{n^2-m^2}\right)
      - \gamma_0^2\right]A.
\eea
Since we have assumed that $T_1$ and $T_2$ are two independent slow
time scales in $A(T_1, T_2)$, we have the following relation when
the original time scale $t$ is used as the temporal parameter for $A(t)$ and $B(t)$:
\bea
\partial_t A = \eta \frac{\partial A}{\partial T_1}
              +\eta^2\frac{\partial A}{\partial T_2},
\\
\partial_t B = \eta \frac{\partial B}{\partial T_1}
              +\eta^2\frac{\partial B}{\partial T_2}.
\eea
We now combine the solvability conditions at ${\cal O}(\eta)$ and at
${\cal O}(\eta^2)$ by using the above two relations, and write both equations 
in the
original time units (also recall that $f=\eta f_0$ and $\gamma = \eta\gamma_0$).
For a frequency ratio $m/n =1/2$, we have
\bea
\partial_t A = \frac{1}{2\omega_0} \left[\frac{f^2}{\omega_0^2}
\left(\frac{3r^2}{2}+\frac{2(1-r)^2}{3}-r(1-r)\cos\phi\right) - \gamma^2\right]B 
\nonumber \\
+ \left(\frac{rf}{\omega_0} 
- \frac{r(1-r)f^2}{2\omega_0^3} \sin\phi - \gamma \right)A,
\\
\partial_t B = - \frac{1}{2\omega_0} \left[\frac{f^2}{\omega_0^2}
\left(\frac{3r^2}{2}+\frac{2(1-r)^2}{3}+r(1-r)\cos\phi\right) - \gamma^2\right]A
\nonumber \\
-\left(\frac{rf}{\omega_0}
- \frac{r(1-r)f^2}{2\omega_0^3} \sin\phi + \gamma \right)B.
\eea
The eigenvalues of the above linear system determine the linear stability
of the flat surface for subharmonic resonance.  The condition for linear 
instability is that at least one eigenvalue has a positive real part.  This 
condition reads,
\bea
\left(\frac{rf}{\omega_0^2}\right)^2 
- \frac{1-r}{r} \left(\frac{rf}{\omega_0^2}\right)^3 \sin\phi 
- \left(\frac{\gamma}{\omega_0}\right)^2 
\nonumber \\
\label{eq:threshold12_1}
> \frac{1}{4}\left[\left(\frac{3}{2}+\frac{2}{3}\left(\frac{1-r}{r}\right)^2\right) 
                    \left(\frac{rf}{\omega_0^2}\right)^2 
                   -\left(\frac{\gamma}{\omega_0}\right)^2
             \right]^2
 -\left(\frac{1-r}{2r}\right)^2\left(\frac{rf}{\omega_0^2}\right)^4
\eea
where we have grouped the dimensionless damping parameter ($\gamma/\omega_0$) 
and forcing amplitude ($rf/\omega_0^2$) together, which are actually the 
expansion parameters.  The RHS of 
the above equation is of higher order in $\gamma/\omega_0$
or in $rf/\omega_0^2$ than the LHS.  However, as we shall see later 
$(1-r)/r \approx 4$ for certain values of $\phi$ along the bicritical
line; thus the first term on the RHS has a quite large coefficient:
$37.0(rf/\omega_0^2)^4$.  Because of that, we do not neglect 
higher order terms on the RHS unless we consider the case 
of extremely 
weak damping $\gamma/\omega_0 \ll 0.16$.  We have also checked that 
corrections to the threshold value of the driving force from orders higher 
than ${\cal O}(\eta^2)$ do {\em not} contain 
terms of the form of $((1-r)/r)^n(rf/\omega_0^2)^n$ for $n>4$.  Therefore 
the truncation at ${\cal O}(\eta^2)$ is a good approximation.

For frequency ratios $m/n \ne 1/2$, the instability condition reads,
\be
\label{eq:thresholdmn_m}
\left(\frac{rf}{m^2\omega_0^2}\right)^2 
- \left(\frac{\gamma}{m\omega_0}\right)^2
> \frac{1}{4}\left[\frac{r^2f^2}{m^4\omega_0^4}
             \left(\frac{3}{2}+\frac{2m^2}{n^2-m^2}\left(\frac{1-r}{r}\right)^2
             \right)-\left(\frac{\gamma}{m\omega_0}\right)^2\right]^2.
\ee

\subsubsection{{$\delta_0 = n^2\omega_0^2$}}

In this case, the condition for the Faraday instability up to 
${\cal O}(\eta)$ reads,
\be
\frac{(1-r)f}{n\omega_0} > \gamma.
\ee
The instability condition up to ${\cal O}(\eta^2)$ for the case of frequency ratio
$m/n=1/2$ is given by,
\bea
\left(\frac{(1\!-\!r)f}{4\omega_0^2}\right) 
+ \left(\frac{2r}{1\!-\!r}\right)^2
  \left(\frac{(1\!-\!r)f}{4\omega_0^2}\right)^3\sin\phi
- \left(\frac{\gamma}{2\omega_0}\right)^2 
+4\left(\frac{r}{1\!-\!r}\right)^4\left(\frac{(1\!-\!r)f}{4\omega_0^2}\right)^4
\nonumber \\
\label{eq:threshold12_2}
> \frac{1}{4}\left[\left(\frac{3}{2}-\frac{2}{3}\left(\frac{2r}{1-r}\right)^2
                   \right) \left(\frac{(1-r)f}{4\omega_0^2}\right)^2 
                   -\left(\frac{\gamma}{2\omega_0}\right)^2
             \right]^2,
\eea
In this case, since $r/(1-r)$ is expected to be small (in M\"uller's
experiment, $r/(1-r) < 0.45$) along the bicritical line, the
approximate threshold value of 
the driving force $f_{th}$ can be found with relatively good accuracy by solving
Eq.~(\ref{eq:threshold12_2}) perturbatively,
\be
\frac{(1-r)f_{th}}{4\omega_0^2} = \frac{\gamma}{2\omega_0} 
+ b_1\left(\frac{\gamma}{2\omega_0}\right)^2
+ b_2 \left(\frac{\gamma}{2\omega_0}\right)^3 + \cdots,
\ee
where
\bea
b_1 = -2\left(\frac{r}{1-r}\right)^2\sin\phi,
\\
b_2= 10 \left(\frac{r}{1-r}\right)^4\sin^2\phi
+ \frac{1}{2}\left[\frac{1}{4}-\frac{4}{3}\left(\frac{r}{1-r}\right)^2\right]^2
- 2\left(\frac{r}{1-r}\right)^4,
\eea
In terms of the driving amplitude $g_{z}$, the threshold value reads,
\be
\label{eq:g_z0_12h}
g_{z}^{12h} = \frac{16\nu k_{12h}\omega_0}{1-r}
        \left[1+b_1\frac{\nu k_{12h}^2}{\omega_0}
               +b_2\left(\frac{\nu k_{12h}^2}{\omega_0}\right)^2\right],
\ee
where $k_{12h}$ is determined by $g_0k_{12h} + \Gamma k_{12h}^3/\rho
= 4\omega_0^2$.  

For frequency ratios that $m/n \ne 1/2$, the instability condition reads,
\be
\label{eq:thresholdmn_n}
\left(\frac{(1-r)f}{n^2\omega_0^2}\right)^2 
- \left(\frac{\gamma}{n\omega_0}\right)^2
> \frac{1}{4}\left[\frac{(1-r)^2f^2}{n^4\omega_0^4}
               \left(\frac{3}{2}-\frac{2n^2}{n^2-m^2}\left(\frac{r}{1-r}\right)^2
               \right) -\left(\frac{\gamma}{n\omega_0}\right)^2\right]^2.
\ee
The threshold value of the driving amplitude $g_{z}$ reads,
\be
\label{eq:g_z0_n}
g_{z}^{(n)} = \frac{8n\nu k_{n}\omega_0}{1-r}
     \left[1 + \frac{1}{8}
           \left(\frac{1}{2}-\frac{2n^2}{n^2-m^2}\left(\frac{r}{1-r}\right)^2
           \right)^2 \left(\frac{2\nu k_{n}^2}{n\omega_0}\right)^2
     \right],
\ee
where $k_{n}$ is determined by $g_0k_{n} + \Gamma k_{n}^3/\rho = n^2\omega_0^2$.

For a frequency ratio $m/n=1/2$, subharmonic resonance and harmonic resonance
will occur simultaneously (for the same value of the driving amplitude) when 
$g_{z}^{(12s)} = g_{z}^{(12h)}$.  The value of $g_{z}^{(12s)}$ is given
by Eq.~(\ref{eq:threshold12_1}), which is written as follows,
\be
\label{eq:threshold12_1_ab}
f_s^2 - \frac{1-r}{r}f_s^3\sin\phi - \gamma_s^2 = 
\frac{1}{4}\left[\left(\frac{3}{2}+\frac{2}{3}\left(\frac{1-r}{r}\right)^2\right)
                f_s^2 - \gamma_s^2\right]^2 
- \left(\frac{1-r}{2r}\right)^2 f_s^2,
\ee
where $f_s = rg_{z}^{(12s)}k_{12s}/(4\omega_0^2)$ and
$\gamma_s = 2\nu k_{12s}^2/\omega_0$.  The value of $g_{z}^{(12h)}$ is given
by Eq.~(\ref{eq:threshold12_2}), which can be written as,
\be
\label{eq:threshold12_2_ab}
f_h^2 + \left(\frac{2r}{1-r}\right)^2 f_h^3 \sin\phi - \gamma_h^2
= \frac{1}{4}\left[\left(\frac{3}{2}-\frac{2}{3}
                                     \left(\frac{2r}{1-r}\right)^2\right)f_h^2
                  -\gamma_h^2\right]^2
  - 4\left(\frac{r f_h}{1-r}\right)^4,
\ee
where $f_h = (1-r)g_{z}^{(12h)}k_{12h}/(16\omega_0^2)$ and 
$\gamma_h = \nu k_{12h}^2/\omega_0$.  The bicritical line $r_b(\phi)$ can be 
obtained directly by solving Eqs.~(\ref{eq:threshold12_1_ab}) and 
(\ref{eq:threshold12_2_ab}) numerically.  Alternatively, $r_b(\phi)$ can be
obtained by substituting the expression of $g_{z}^{(12h)}$ of 
Eq.~(\ref{eq:g_z0_12h}) as well as that of $g_{z}^{(12s)}$ into 
Eq.~(\ref{eq:threshold12_1_ab}), and then solve Eq.~(\ref{eq:threshold12_1_ab}).
Except for a small difference near $\phi=\pi/2$, where $r$ is largest on the
bicritical line, the second way is much easier.  In the results
presented below, the values of $r_b(\phi)$ directly calculated from
Eq.~(\ref{eq:threshold12_1_ab}) and (\ref{eq:threshold12_2_ab}) are used.

\begin{figure}[t]
\hspace{-3cm}\psfig{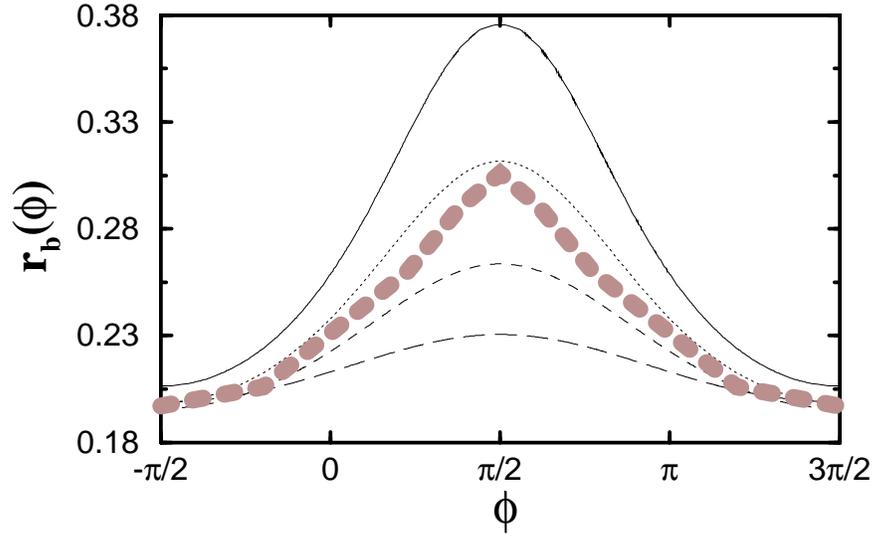}
\vspace{-3cm}
\caption{Bicritical lines of Faraday waves driven by a two-frequency force 
of frequency ratio $m/n= 1/2$.  The gray symbols are taken from M\"uller's
experiment (M\"uller 1993), and the solid curve is calculated using
parameters of M\"uller's experiment: $\nu=0.2 \mbox{cm}^2/\mbox{s}$, 
$\omega_0/2\pi=27.9 \mbox{Hz}$, $k_{12s}=8.7 \mbox{cm}^{-1}$, and 
$k_{12h}=17.0 \mbox{cm}^{-1}$.   Other curves are calculated using the same 
values of $\omega_0$, $k_{12s}$, and $k_{12h}$ as the solid curve, but 
$\nu=0.15 \mbox{cm}^2/\mbox{s}$ for the dotted curve, 
$\nu=0.1 \mbox{cm}^2/\mbox{s}$ for the dashed curve, and 
$\nu=0.05 \mbox{cm}^2/\mbox{s}$ for the long-dashed curve.}
\label{fig-Muller_bline}
\end{figure}

The solid curve in Figure \ref{fig-Muller_bline} 
is the bicritical line calculated by using parameters appropriate for the 
experimental data of \cite{re:muller93}.  Note that for simplicity we 
have considered surface waves in the infinite depth limit in the above 
calculations.  By assuming that the viscous dissipation near the bottom 
boundary layer can be neglected, our results can be generalized to the 
case of a fluid layer of finite depth $d$ by replacing the infinite depth 
dispersion relation by $\omega(k)^2 = \tanh(kd)(g_0k+\Gamma k^3/\rho)$ and
$g_{z}$ by $g_{z}\tanh(kd)$.  Since $k_{12s}d=2.0$ and $k_{12h}d=3.9$ in
M\"uller's experiments, such finite depth corrections are small, but have been
included in Fig.~\ref{fig-Muller_bline}. We also note that the value of 
$k_{12s}$ calculated from the finite depth dispersion relation using appropriate
fluid and forcing parameters is different from that observed 
experimentally by M\"uller,
and we have used the experimental value of wavenumber $k_{12s}$ 
\footnote{The value of $k_{12s}$ calculated from the finite depth dispersion 
relation using appropriate fluid and forcing parameters 
($\rho=0.95 > \mbox{g/cm}^3$, $\Gamma=20.6 \> \mbox{dyn/cm}$, 
$d=0.23 \> \mbox{cm}$, 
and $\omega_0 = 2\pi\times 27.9 \> \mbox{Hz}$) is $10 \> \mbox{cm}^{-1}$, or
$\lambda_{12s}=0.63 \> \mbox{cm}$.  This value of the critical wavelength is
significantly different from what was observed by M\"uller, 
$\lambda_{12s} \approx 0.72 \> \mbox{cm}$.  The calculated value of 
$k_{12h}$ does agree well with the observed 
value, $k_{12h}=17 \mbox{cm}^{-1}$, or $\lambda_{12h} \approx 0.37 \mbox{cm}$.}.
Our analytical results agree
qualitatively with the experimental results of M\"uller, shown as gray symbols.
Other curves in Fig.~\ref{fig-Muller_bline} are bicritical lines for the same
values of $\omega_0$, $k_{12s}$, and $k_{12h}$ as the solid curve, but
$\nu=0.15$ for the dotted curve, $\nu=0.1$ for the dashed curve, and $\nu=0.05$ 
for the long-dashed curve. Therefore our calculation is in qualitative 
agreement with the bicritical line determined experimentally. The small
quantitative discrepancy between our analytical results and experimental 
results is probably due to the relatively large values of the damping parameters 
($\gamma_s=0.17$ and $\gamma_h=0.33$). It is likely that
the quasi-potential approximation is not accurate in this parameter range.

For frequency ratios $m/n \ne 1/2$, we predict that the bicritical line is 
independent of the phase difference $\phi$, and can be readily
obtained from Eq.~(\ref{eq:thresholdmn_m}) and (\ref{eq:thresholdmn_n}).  
Further experimental studies of 
bicritical lines for different values of frequency ratio, $\nu$, $\omega_0$, 
and surface tension $\Gamma$ would be interesting to provide additional tests 
of our results.  

\section{Standing Wave Amplitude Equations}
\label{sec:SWAE_two_frequency}

We present in this section a weakly nonlinear analysis of parametric
surface waves driven by two frequencies, extending the calculations
presented earlier for the single frequency case by \cite{re:zhang96b}.
For fluids of low viscosity, the equations governing fluid flow and the
boundary conditions at the free surface were expanded in the (small)
width of the vortical layer adjacent to the free surface. The bulk flow
is then potential, but satisfies effective boundary conditions on the
moving surface. We furthermore neglected viscous terms that are
nonlinear in the free surface variables to arrive at the so-called
Linear Damping Quasi-Potential equations. They involve only
the surface's deviation away from planarity $h$, and the surface
velocity potential $\Phi$, but no longer depend on the bulk velocity
field,
\bea
\label{eq:hmn}
\partial_t h(\vx,t)  = \gamma\nabla^2 h + \sD\Phi - \nabla \cdot (h\nabla\Phi)
    + \frac{1}{2} \nabla^2 (h^2\sD\Phi)
\nonumber \\
 -\sD(h\sD\Phi) + \sD\left[h\sD(h\sD\Phi) + \frac{1}{2}h^2\nabla^2\Phi\right],
\\
\label{eq:Phimn}
\partial_t \Phi(\vx,t) = \gamma\nabla^2\Phi
                       - \left(G_0-\Gamma_0\nabla^2\right) h 
                       - 4f \biggl(\sin 2t + \kappa \sin (2pt + \Theta)\biggr)h
\nonumber \\
     + \frac{1}{2}\left(\sD\Phi\right)^2
     - \frac{1}{2}\left(\nabla\Phi\right)^2
     -(\sD\Phi)\left[h\nabla^2\Phi + \sD(h\sD\Phi)\right]
     -\frac{\Gamma_0}{2}\nabla\cdot\left(\nabla h(\nabla h)^2\right).
\eea
Linearization of these two equations leads to Eq. (\ref{eq:h2_12_}).
When the forcing 
component $\sin 2m\omega_0t$ dominates, we choose $1/(m\omega_0)$ as the unit 
of time and $1/k_{0m}$ as the unit of length with $g_0k_{0m} 
+ \Gamma k_{0m}^3/\rho = m^2\omega_0^2$.  For the other case when the forcing 
component $\sin 2n\omega_0t$ dominates, we choose $1/(n\omega_0)$ as the unit 
of time and $1/k_{0n}$ as the unit of length with $g_0k_{0n} 
+ \Gamma k_{0n}^3/\rho = n^2\omega_0^2$.  $\sD$ is a linear and nonlocal
operator that multiplies each Fourier component of a field by its
wavenumber modulus (Zhang \& Vi\~nals 1996b).
We can write the system of equations for the two cases in the same 
dimensionless form,
with the dimensionless variables $\gamma$, $G_0$, $\Gamma_0$, $f$, $\kappa$, 
and $p$ having different values for the two cases.  When the forcing
component $\sin 2m\omega_0t$ dominates, we have
\bea
\gamma = \frac{2\nu k_{0m}^2}{m\omega_0}, \>\>\>\>\>\> 
     f = \frac{g_{z}k_{0m}r}{4m^2\omega_0^2}, \>\>\>\>\>\>
 \kappa= \frac{1-r}{r}, \>\>\>\>\>\> p = \frac{n}{m},
\\ 
     G_0 = \frac{g_0k_{0m}}{m^2\omega_0^2}, \>\>\>\>
\Gamma_0 = \frac{\Gamma k_{0m}^3}{\rho m^2\omega_0^2}, \>\>\>\>
    (G_0 + \Gamma_0 = 1), \>\>\>\>\>\> \Theta = \phi.
\eea
For the other case (the forcing component $\sin 2n\omega_0t$ dominates), we
have
\bea
\gamma = \frac{2\nu k_{0n}^2}{n\omega_0}, \>\>\>\>\>\> 
     f = \frac{g_{z}k_{0n}(1-r)}{4n^2\omega_0^2}, \>\>\>\>\>\>
 \kappa= \frac{r}{1-r}, \>\>\>\>\>\> p = \frac{m}{n},
\\
     G_0 = \frac{g_0k_{0n}}{n^2\omega_0^2}, \>\>\>\>
\Gamma_0 = \frac{\Gamma k_{0n}^3}{\rho n^2\omega_0^2}, \>\>\>\> 
    (G_0 + \Gamma_0 = 1), \>\>\>\>\>\> \Theta = -\frac{m}{n}\phi.
\eea

\subsection{Derivation of SWAE's}
\label{sec:derivation_SWAE}

As mentioned above, the detailed procedure for the derivation of
standing wave amplitude 
equations parallels that presented by \cite{re:zhang96b} for the case of 
sinusoidal forcing.  We seek nonlinear 
standing wave solutions of Faraday waves near onset, i.e. 
$\varepsilon = (f-\gamma)/\gamma \ll 1$.  The quasi-potential
equations (Eqs.~(\ref{eq:hmn}) and (\ref{eq:Phimn})) are expanded consistently
in $\varepsilon^{1/2}$ with multiple time scales,
\bea
\label{eq:expansion_h_two}
h(\vx, t, T) = \varepsilon^{1/2} h_1(\vx, t, T) + \varepsilon h_2
             + \varepsilon^{3/2} h_3 + \cdots, \\
\label{eq:expansion_Phi_two}
\Phi(\vx, t, T) = \varepsilon^{1/2} \Phi_1(\vx, t, T) + \varepsilon \Phi_2
             + \varepsilon^{3/2} \Phi_3 + \cdots,
\eea
where $h_1$ and $\Phi_1$ are the linear neutral solutions and can be
found in a similar way to the case of sinusoidal forcing.  For
simplicity, we shall only consider terms up to order $f$ or $\gamma$ 
for $h_1$ and $\Phi_1$.  They are
\bea
h_1(\vx,t) = \Biggl[\cos t + \frac{\gamma}{4}\sin 3t 
         + \frac{\gamma \kappa}{2p(p+1)} \sin\biggl((2p+1)t+\Theta\biggr)
\nonumber \\
         + \frac{\gamma \kappa}{2p(p-1)} \sin\biggl((2p-1)t+\Theta\biggr) \Biggr]
         \sum_{j=1}^{N}\left[A_j(T)\exp\left(\hkj\cdot\vx\right) + c.c \right],
\\
\Phi_1(\vx,t) = \Biggl[-\sin t + \gamma \cos t + \frac{3\gamma}{4}\cos 3t
      + \frac{\gamma \kappa(2p+1)}{2p(p+1)} \cos\biggl((2p+1)t+\Theta\biggr)
\nonumber \\
      + \frac{\gamma\kappa(2p-1)}{2p(p-1)} \cos\biggl((2p-1)t+\Theta\biggr)\Biggr]
      \sum_{j=1}^{N}\left[A_j(T)\exp\left(\hkj\cdot\vx\right) + c.c \right],
\eea 
where $T=\varepsilon t$.

At ${\cal O}(\varepsilon)$, we have the following 
non-homogeneous linear equation for $h_2$,
\bea
\partial_{tt}h_2 - 2\gamma\nabla^2\partial_t h_2
+ (G_0-\Gamma_0\nabla^2)\sD h_2 
+ 4\gamma \biggl(\sin 2t + \kappa\sin(2pt+\Theta) \biggr) \sD h_2
\nonumber \\
= \sum_{j,l=1}^{N}\Biggl\{\frac{1+c_{jl}}{4}\sqrt{2(1+c_{jl})}
  -\cos 2t \left[1+c_{jl}-\frac{3-c_{jl}}{4}\sqrt{2(1+c_{jl})}\right]
\nonumber \\
  -\gamma\sin 2t \left[(5/2+c_{jl})\left(1+c_{jl}-\sqrt{2(1+c_{jl})}\right)
  +\frac{1+c_{jl}}{8}\sqrt{2(1+c_{jl})}\right]
\nonumber \\
+\frac{2\gamma\kappa}{1-p^2}\sin(2pt+\Theta)
 \left[\frac{1+c_{jl}}{4}\sqrt{2(1+c_{jl})} +
       p^2\left(1+c_{jl}-\sqrt{2(1+c_{jl})}\right)\right]
\nonumber \\
-\frac{\gamma\kappa}{p}\sin[2(p\!+\!1)t\!+\!\Theta]
 \left[\frac{2p+1}{p+1}(1\!+\!c_{jl})\sqrt{2(1\!+\!c_{jl})} +
 (p\!+\!1)\left(1\!+\!c_{jl}\!-\!\sqrt{2(1\!+\!c_{jl})}\right)\right]
\nonumber \\
+\frac{\gamma\kappa}{p}\sin[2(p\!-\!1)t\!+\!\Theta]
 \left[\frac{2p\!-\!1}{p\!-\!1}(1\!+\!c_{jl})\sqrt{2(1\!+\!c_{jl})} +
 (p\!-\!1)\left(1\!+\!c_{jl}\!-\!\sqrt{2(1\!+\!c_{jl})}\right)\right] \Biggr\}
\nonumber \\
 \times \left[A_jA_l\exp\left(i(\hkj+\hkl)\cdot\vx\right) + c.c.\right]
\nonumber \\
+ \sum_{j,l=1}^{N}\Biggl\{\frac{1-c_{jl}}{4}\sqrt{2(1-c_{jl})}
  -\cos 2t \left[1-c_{jl}-\frac{3+c_{jl}}{4}\sqrt{2(1-c_{jl})}\right]
\nonumber \\
  -\gamma\sin 2t \left[(5/2-c_{jl})\left(1-c_{jl}-\sqrt{2(1-c_{jl})}\right)
  +\frac{1-c_{jl}}{8}\sqrt{2(1-c_{jl})}\right]
\nonumber \\
+\frac{2\gamma\kappa}{1-p^2}\sin(2pt+\Theta)
 \left[\frac{1-c_{jl}}{4}\sqrt{2(1-c_{jl})} +
       p^2\left(1-c_{jl}-\sqrt{2(1-c_{jl})}\right)\right]
\nonumber \\
-\frac{\gamma\kappa}{p}\sin[2(p\!+\!1)t\!+\!\Theta]
 \left[\frac{2p+1}{p+1}(1\!-\!c_{jl})\sqrt{2(1\!-\!c_{jl})} +
 (p\!+\!1)\left(1\!-\!c_{jl}\!-\!\sqrt{2(1\!-\!c_{jl})}\right)\right]
\nonumber \\
+\frac{\gamma\kappa}{p}\sin[2(p\!-\!1)t+\Theta]
 \left[\frac{2p\!-\!1}{p\!-\!1}(1\!-\!c_{jl})\sqrt{2(1\!-\!c_{jl})} +
 (p\!-\!1)\left(1\!-\!c_{jl}\!-\!\sqrt{2(1\!-\!c_{jl})}\right)\right] \Biggr\}
\nonumber \\
\label{eq:SWAE_2f_order2}
\!\!\!\!\!\!\!\!\!\!\!\!\!\!\!\!\!\!\!\!\!\!\!\!\!\!\!\!\!\!\!\!\!\!\!\!\!
\times\left[A_jA_l^*\exp\left(i(\hkj\!-\!\hkl)\cdot\vx\right) + c.c.\right].
 \>\>\>\>\>
\eea

Before obtaining the detailed form of the equations, let us mention that
the generic form of the SWAE's for two-frequency driven Faraday waves can 
be obtained by symmetry considerations ( Edwards \& Fauve 1994).  For 
the case of single frequency forcing the SWAE's derived in \cite{re:zhang96b} 
contain only third order nonlinear terms, but no 
quadratic terms.   The exclusion of quadratic terms can be understood there from
the requirement of sign invariance of the SWAEs.   Subharmonic response of
the fluid surface to the driving force $f\sin(2\omega_0 t)$ implies
$h_j(\vx, t+\pi/\omega_0) = - h_j(\vx, t)$.   Here $h_j$ is a linear unstable
mode,
\be
h_j = \left(\cos \omega_0 t + \frac{f}{4}\sin 3\omega_0 t + \cdots) \right)
      \left(A_j\exp\left(i\hkj\cdot\vx\right) + c.c.\right),
\ee
where only odd multiples of frequency $\omega_0$ appear.  As a result,
a sign change of $A_j$ is equivalent to a time displacement in a period of
the driving force, $t \rightarrow t+\pi/\omega_0$.  Because of the invariance 
of the surface wave system under such a time displacement, the amplitude
equation of $A_j$ must be sign invariant, which obviously excludes quadratic
terms.   However, quadratic terms can arise with two-frequency forcing. 
If the frequency ratio 
$m/n =$ even/odd or odd/even, the SWAE's are sign invariant if the odd 
frequency dominates, and otherwise if the even frequency dominates.  
When the frequency
ratio $m/n =$ odd/odd, the SWAE's are sign invariant. A general consequence
of the loss of sign invariance is the appearance of quadratic terms in the 
amplitude equation. 

We concentrate here on the case of frequency ratio $m/n=1/2$, since detailed
experimental results are available for this case.  Similar calculations can
be done for other frequency ratios, and they might be interesting for future
experimental studies.  We further restrict our attention to the case
that the odd frequency dominates, which also corresponds to the
subharmonic side of the bicritical line.  In this case, 
the solutions at ${\cal O}(\varepsilon^{1/2})$ are,
\bea
h_1 \!=\! \left[\cos t + \frac{\gamma}{4}\sin 3t 
           +\frac{\gamma\kappa}{4}\sin(3t+\phi) \right]
      \sum_{j=1}^{N}\left[A_j\exp\left(\hkj\!\cdot\!\vx\right) \!+\! c.c \right],
\\
\Phi_1 \!=\! \left[-\sin t + \gamma \cos t + \frac{3\gamma}{4}\cos 3t
               +\frac{3\gamma\kappa}{4}\cos (3t\!+\!\phi) \right]
          \sum_{j=1}^{N}\left[A_j\exp\left(\hkj\!\cdot\!\vx\right)\!+\!c.c \right],
\eea
where $\kappa=(1-r)/r$.  At ${\cal O}(\varepsilon)$, we have,
\bea
\partial_{tt}h_2 - 2\gamma\nabla^2\partial_t h_2
+ (G_0-\Gamma_0\nabla^2)\sD h_2
+ 4\gamma \biggl(\sin 2t + \kappa\sin(4t+\phi) \biggr) \sD h_2
\nonumber \\
= \sum_{j,l=1}^{N}\Biggl\{\frac{1+c_{jl}}{4}\sqrt{2(1+c_{jl})}
  -\cos 2t \Biggl[1+c_{jl}-\frac{3-c_{jl}}{4}\sqrt{2(1+c_{jl})}
\nonumber \\
                -\frac{\gamma\kappa}{2}\sin\phi\left(1+c_jl+(2+3c_{jl})
                                                    \sqrt{2(1+c_{jl})}\right)
           \Biggl]
\nonumber \\
  -\gamma\sin 2t \Biggl[(5/2+c_{jl})\left(1+c_{jl}-\sqrt{2(1+c_{jl})}\right)
                       +\frac{1+c_{jl}}{8}\sqrt{2(1+c_{jl})} 
\nonumber \\
                      -\frac{\kappa}{2}\cos\phi\left(1+c_jl+(2+3c_{jl})
                                                    \sqrt{2(1+c_{jl})}\right)
                 \Biggl]\Biggr\}
  \left[A_jA_l\exp\left(i(\hkj+\hkl)\cdot\vx\right) + c.c.\right]
\nonumber \\
+ \sum_{j,l=1}^{N}\Biggl\{\frac{1-c_{jl}}{4}\sqrt{2(1-c_{jl})}
  -\cos 2t \Biggl[1-c_{jl}-\frac{3+c_{jl}}{4}\sqrt{2(1-c_{jl})}
\nonumber \\
                -\frac{\gamma\kappa}{2}\sin\phi\left(1-c_{jl}+(2-3c_{jl})
                                                    \sqrt{2(1-c_{jl})}\right)
           \Biggl]
\nonumber \\
  -\gamma\sin 2t \Biggl[(5/2-c_{jl})\left(1-c_{jl}-\sqrt{2(1-c_{jl})}\right)
                       +\frac{1-c_{jl}}{8}\sqrt{2(1-c_{jl})}
\nonumber \\
                      -\frac{\kappa}{2}\cos\phi\left(1-c_{jl}+(2-3c_{jl})
                                                    \sqrt{2(1-c_{jl})}\right)
                 \Biggl]\Biggr\}
 \left[A_jA_l^*\exp\left(i(\hkj-\hkl)\cdot\vx\right) + c.c.\right].
\eea

We now solve the above equation for $h_2$. 
Since there are no secular terms on the RHS, we are only
interested
in the particular solution for $h_2$ caused by the RHS of that equation.
The particular solution reads,
\bea
h_2 = \sum_{j,l=1}^{N}\biggl\{
    \left(\alpha_{jl}+\beta_{jl}\cos 2t+\gamma\delta_{jl}\sin 2t\right)
    \left[A_jA_l\exp\left(i(\hkj+\hkl)\cdot\vx\right) + c.c.\right]
\nonumber \\
\label{eq:h_order2_SWAE_2f}
   +\left(\bar{\alpha}_{jl}+\bar{\beta}_{jl} \cos 2t +
     \gamma\bar{\delta}_{jl} \sin 2t\right)
    \left[A_jA_l^*\exp\left(i(\hkj-\hkl)\cdot\vx\right) + c.c.\right]\biggr\},
\eea
where
\bea
\alpha_{jl} = \frac{1+c_{jl}}{4\left[G_0+2\Gamma_0(1+c_{jl})\right]}
            - \frac{2\gamma^2 \delta_{jl}}{G_0 + 2\Gamma_0 (1+c_{jl})},
\\
\beta_{jl}  = 
\frac{-H_{jl}\left(E_{jl}-8\gamma^2M_{jl}\right)
      +\gamma^2\left(8(1+c_{jl})-2\kappa\sqrt{2(1+c_{jl})}\cos\phi\right)N_{jl} }
     {\gamma^2\left(8(1+c_{jl})-2\kappa\sqrt{2(1+c_{jl})}\cos\phi\right)^2
      + E_{jl}^2 - 8\gamma^2 E_{jl}M_{jl} },
\\
\delta_{jl} =
-\frac{\left(8(1+c_{jl})-2\kappa\sqrt{2(1+c_{jl})}\cos\phi\right)H_{jl}
       +E_{jl}N_{jl} }
     {\gamma^2\left(8(1+c_{jl})-2\kappa\sqrt{2(1+c_{jl})}\cos\phi\right)^2
      +E_{jl}^2 -8\gamma^2 E_{jl}M_{jl}},\\
\label{eq:M_jl}
M_{jl} = \frac{\sqrt{2(1+c_{jl})}}{G_0+2\Gamma_0(1+c_{jl})}, \\
\label{eq:D_jl}
D_{jl} = \left[G_0+2\Gamma_0(1+c_{jl})\right]\sqrt{2(1+c_{jl})} - 4,
\label{eq:N_jl}
\eea
\be
\label{eq:eij}
E_{jl} = D_{jl}+2\kappa\gamma\sqrt{2(1+c_{jl})}\sin\phi,
\ee
\be
H_{jl} = 1+c_{jl}-\frac{3-c_{jl}}{4}\sqrt{2(1+c_{jl})}
        -\frac{\kappa\gamma}{2}\left(1+c_{jl}+(2+3c_{jl})\sqrt{2(1+c_{jl})}\right)
         \sin\phi,
\ee
and 
\bea
N_{jl} = \left(\frac{5}{2}+c_{jl}\right)\left(1+c_{jl}-\sqrt{2(1+c_{jl})}\right)
      + \frac{1+c_{jl}}{8}\sqrt{2(1+c_{jl})} + (1+c_{jl})M_{jl}
\nonumber \\
      - \frac{\kappa}{2}\left(1+c_{jl}+(2+3c_{jl})\sqrt{2(1+c_{jl})}\right)
        \cos\phi.
\eea
$\bar{\alpha}_{jl}$, $\bar{\beta}_{jl}$, and $\bar{\delta}_{jl}$ can be
obtained by replacing $c_{jl}$ with $-c_{jl}$ in the expressions for
$\alpha_{jl}$, $\beta_{jl}$, and $\delta_{jl}$ respectively.

The solution for $\Phi_2$ is,
\bea
\Phi_2 = \sum_{j,l=1}^{N}\biggl\{
    \left(\gamma u_{jl}+\gamma v_{jl}\cos 2t + w_{jl}\sin 2t\right)
    \left[A_jA_l\exp\left(i(\hkj+\hkl)\cdot\vx\right) + c.c.\right]
\nonumber \\
   +\left(\gamma\bar{u}_{jl}+\gamma\bar{v}_{jl}\cos 2t+\bar{w}_{jl}\sin
2t\right)
    \left[A_j A_l^*\exp\left(i(\hkj-\hkl)\cdot\vx\right) + c.c.\right]\biggr\},
\eea
where
\bea
u_{jl} = \frac{1}{2} + \left(\alpha_{jl}-\frac{1}{4}\right)\sqrt{2(1+c_{jl})},
\\
\label{eq:vjl_2f}
v_{jl} = \frac{3}{4} + \left(\beta_{jl}-\frac{3}{8}\right)\sqrt{2(1+c_{jl})}
       + \frac{2\delta_{jl}}{\sqrt{2(1+c_{jl})}},
\\
\label{eq:wjl_2f}
w_{jl} = -\frac{1}{2} + \frac{1}{4}\sqrt{2(1+c_{jl})}
       - \frac{2\beta_{jl}}{\sqrt{2(1+c_{jl})}},
\eea
and $\bar{u}_{jl}$, $\bar{v}_{jl}$, and $\bar{w}_{jl}$ can be
obtained by replacing $c_{jl}$ with $-c_{jl}$ in the expressions for
$u_{jl}$, $v_{jl}$, and $w_{jl}$ respectively.       

\begin{figure}[t]
\hspace{-1.5cm}\psfig{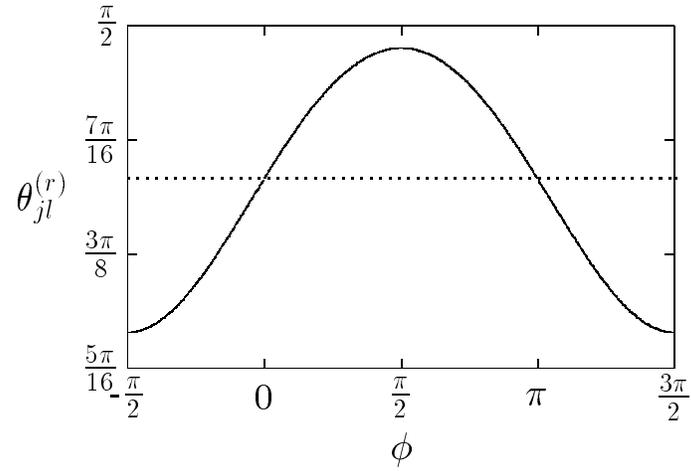}
\vspace{-7cm}
\caption{Modified triad resonant condition, $\theta_{jl}^{(r)}$ as a function of
the phase difference $\phi$, for the subharmonic response in
two-frequency Faraday waves of frequency ratio 1:2 with $\Gamma_0=1$,
$\gamma=0.15$, and $r=0.3$.  The dashed line corresponds to the value of
$\theta_{jl}^{(r)}$ for purely capillary waves in the case of sinusoidal
forcing.}
\label{fig-tp_15_G10_30}
\end{figure}

We note that the triad resonant condition is $E_{jl}=0$ in 
Eq. (\ref{eq:eij}), which is different
from that of the case of sinusoidal forcing ($D_{jl}=0$) 
(Zhang \& Vi\~nals 1996b). This difference
is important since it results in different values of $c_{jl}$ at the triad
resonance, which is now also a function of the damping parameter
$\gamma$, the ratio of amplitudes $r$, and
the phase difference $\phi$.  As an example, Figure \ref{fig-tp_15_G10_30}
shows the modified triad resonant condition for $\Gamma_0=1$, $\gamma=0.15$, 
and $r=0.3$.  As we see that the value of $\theta_{jl}^{(r)}$ is close to
$90^{\circ}$ for some values of $\phi$.  
From the results in the case of sinusoidal forcing, we know that
when the wavevectors of two standing waves are separated by an angle
of $\theta_{jl}^{(r)}$, the pattern is strongly suppressed.
Two waves of wavevectors separated by this angle will excite a linearly
stable mode with an amplitude inversely proportional to the damping
coefficient. Energy is then dissipated by the stable mode, and hence
the two original waves are effectively damped as compared to other
unstable modes that do not satisfy the triad resonance condition.
Thus, without further calculations one 
already can conclude that square patterns are not favored when the
phase difference $\phi$ 
is close to $\pi/2$ (or $\varphi$ close to 0 in M\"uller's notation)
for this set of parameters.  

At ${\cal O}(\varepsilon^{3/2})$, the calculation is again similar to the
case of single frequency forcing.  The solvability condition gives the following
standing wave amplitude equations
\be
\label{eq:SWA_two1}
\frac{1}{\gamma}\frac{\partial A_j}{\partial T} =  A_j - \left[g(1)|A_j|^2
              +\sum_{l=1(l\ne j)}^{N} g(c_{jl})|A_l|^2\right] A_j,
\ee
where $j=1,2, \cdots, N$, 
\be
g(1) = \frac{28+9\Gamma_0 + (12+9\Gamma_0)\kappa\cos\phi}{64} +
2\alpha_{jj}+\frac{3(1+\kappa\cos\phi)}{8}\beta_{jj}-\frac{1}{2}\delta_{jj},
\ee
and 
\bea
g(c_{jl}) = \frac{3\Gamma_0(1\!+\!\kappa\cos\phi)}{32}\left(1\!+\!2c_{jl}^2\right)
\!+\! \frac{7\!+\!3\kappa\cos\phi}{8}
\left(3\!-\!\sqrt{2(1\!+\!c_{jl})}\!-\!\sqrt{2(1\!-\!c_{jl})}\right)
\nonumber \\
          + \left(1+c_{jl}-\sqrt{2(1+c_{jl})}\right)
            \left(\frac{1+\kappa\cos\phi}{4}w_{jl} - v_{jl}\right)
\nonumber \\
          + \left(1-c_{jl}-\sqrt{2(1-c_{jl})}\right)
            \left(\frac{1+\kappa\cos\phi}{4}\bar{w}_{jl} - \bar{v}_{jl}\right)
\nonumber \\
    + (1+c_{jl})\left(2\alpha_{jl}
      + \frac{3(1+\kappa\cos\phi)}{8}\beta_{jl} - \frac{1}{2}\delta_{jl}\right)
\nonumber \\
    + (1-c_{jl})\left(2\bar{\alpha}_{jl}
      + \frac{3(1+\kappa\cos\phi)}{8}\bar{\beta}_{jl}
                        - \frac{1}{2}\bar{\delta}_{jl}\right).
\eea 
When $\kappa=0$, Eq.~(\ref{eq:SWA_two1}) reduces to the standing wave amplitude
equation for sinusoidal forcing as expected (Zhang \& Vi\~nals 1996b).

In contrast to single frequency forced Faraday waves, we note that
$g(1)<0$ for some ranges of parameter values in this two frequency case.
When $g(1)<0$, the amplitude $A_j$ would increase without limit.
Therefore higher order terms (at least of fifth order) will be
required in the SWAE to saturate the amplitude.  The regions
where $g(1)<0$ for particular sets of parameters are shown as black
in Fig.~\ref{fig-r_phi_2f_selection}.
The steady state of parametric surface waves in parameter regions 
such that $g(1)<0$ cannot be determined from the SWAEs 
(Eq.~(\ref{eq:SWA_two1})).  Analytical calculations of the relevant 
higher order terms in this case are algebraically much more complicated 
than the calculations presented here, and we have not done such 
calculations.  We shall restrict ourselves to the parameter range 
that gives a positive $g(1)$ in further analysis of this section.  

When $g(1)>0$, we can rescale the
amplitude $A_j$ in Eq.~(\ref{eq:SWA_two1}) as $\tilde{A}_j =
\sqrt{g(1)}A_j$. 
We have the
following standing wave amplitude equation for the scaled amplitude,
\be
\label{eq:SWAE_scaled}
\frac{1}{\gamma}\frac{\partial \tilde{A}_j}{\partial T} =  \tilde{A}_j
- \left[|\tilde{A}_j|^2 +\sum_{l=1(l\ne j)}^{N} \tilde{g}(c_{jl})
|\tilde{A}_l|^2 \right] \tilde{A}_j,
\ee
where $\tilde{g}(c_{jl})=g(c_{jl})/g(1)$. Note that
$\tilde{g}(c_{jl} \rightarrow \pm 1) = 2$.

\begin{figure}[t]
\hspace{-2.0cm}
\psfig{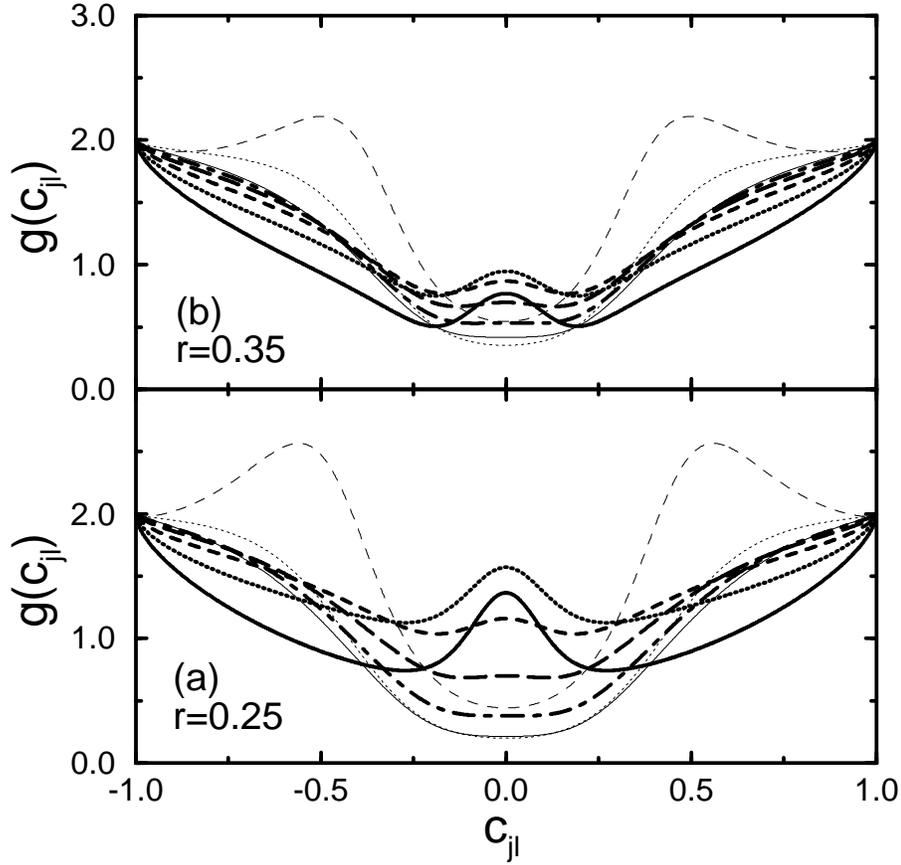}
\caption{Nonlinear coefficient $g(c_{jl})$ of the SWAEs for two-frequency driven
Faraday waves for purely capillary waves with damping parameter $\gamma=0.1$ and
the relative amplitude $r=0.25$ in (a), $r=0.35$ in (b).  The phase difference
$\phi=\pi/2$ for the thick solid curves, $\phi=0.6\pi$ for the thick dotted
curves, $\phi=0.7\pi$ for the thick dashed curves, $\phi=0.8\pi$ for the thick 
long-dashed curves, $\phi=0.9\pi$ for the thick dot-dashed curves, $\phi=\pi$
for the thin solid curves, $\phi=1.3\pi$ for the thin dotted curves, and
$\phi=3\pi/2$ for the thin dashed curves.}
\label{fig-gc_100_10}
\end{figure}

The scaled nonlinear interaction coefficient $g(c_{jl})$ 
(we have suppressed the tilde since we will only refer to the scaled nonlinear
coefficient in what follows) for the regions with a positive $g(1)$ 
depends strongly on the phase
difference $\phi$.  This is due to the effect of the modified triad
resonant condition (see Fig.~\ref{fig-tp_15_G10_30}), as well as other effects,
such as the dependence of the surface wave amplitude at the modified triad 
resonance on $\phi$.  As an example, Fig.~\ref{fig-gc_100_10}(a) 
shows the scaled coefficient $g(c_{jl})$ as a function of $c_{jl}$ for
purely capillary waves with
$\gamma=0.1$ and $r=0.25$.  The general trend in these curves can be understood
from the different triad resonant condition for different values of $\phi$.  
As $\phi$ decreases from $3\pi/2$ to $\pi/2$,  the angle $\theta_{jl}^{(r)}$ 
(recall that $c_{jl} = \cos(\theta_{jl})$) at the triad resonance increases to 
a value close to $\pi/2$ (see Fig.~\ref{fig-tp_15_G10_30}), and therefore
the values of $g(c_{jl})$ changes from a minimum ($<1$) to a local maximum 
($>1$).  This change in $g(c_{jl})$
makes a pattern of square symmetry unstable for $\phi$ close to $\pi/2$ 
(since $g(0) > 1$).  
By increasing the values of $r$ in Fig.~\ref{fig-gc_100_10}(b), 
the amplitude of the smaller forcing component ($\kappa\sin(4t+\phi)$) 
decreases, and therefore the changes in $g(c_{jl})$ for different values of
$\phi$ are smaller.  
We note that with the damping parameters used in
Fig.~\ref{fig-gc_100_10}, surface waves may have synchronous (harmonic)
response to the driving force, i.e. below the bicritical line in 
Fig.~\ref{fig-Muller_bline} for some values of the phase difference
$\phi$.  In that case, the above results of $g(c_{jl})$ become
irrelevant for these values of $\phi$.
 
\begin{figure}[t]
\hspace{-2.0cm}
\psfig{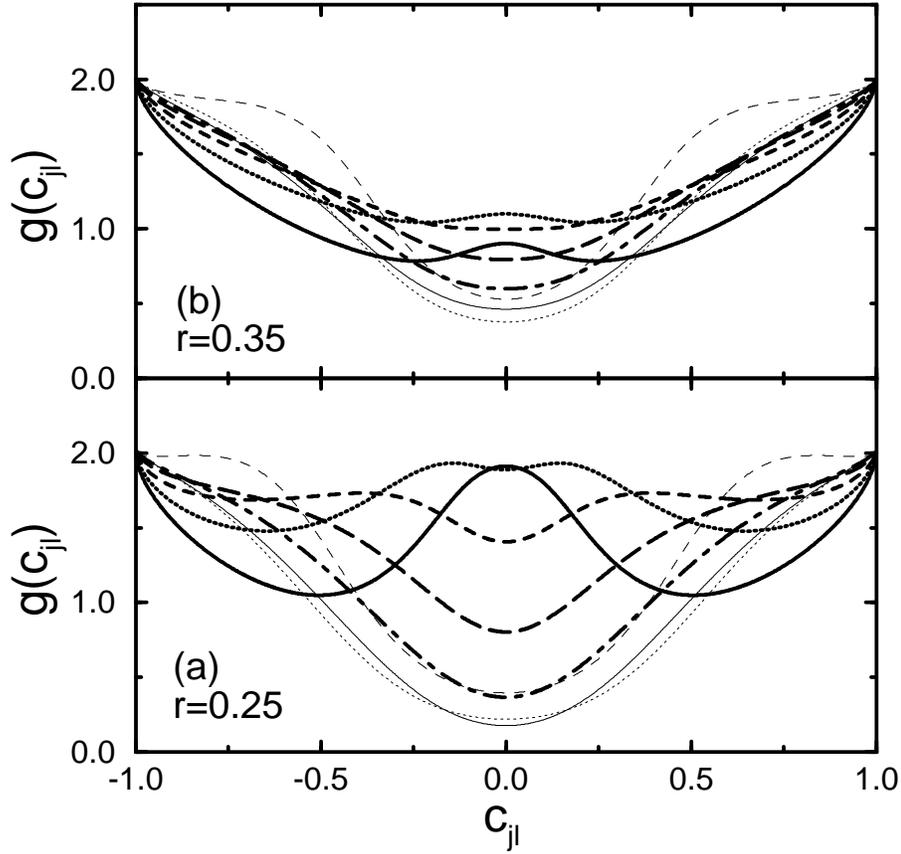}
\caption{Nonlinear coefficient $g(c_{jl})$ of the SWAEs for two-frequency driven
Faraday waves for purely capillary waves with damping parameter 
$\gamma=0.15$ and
the relative amplitude $r=0.25$ in (a), $r=0.35$ in (b).  The phase difference
$\phi=\pi/2$ for the thick solid curves, $\phi=0.6\pi$ for the thick dotted
curves, $\phi=0.7\pi$ for the thick dashed curves, $\phi=0.8\pi$ for the thick
long-dashed curves, $\phi=0.9\pi$ for the thick dot-dashed curves, $\phi=\pi$
for the thin solid curves, $\phi=1.3\pi$ for the thin dotted curves, and
$\phi=3\pi/2$ for the thin dashed curves.}
\label{fig-gc_100_15}
\end{figure}

For larger values of the damping parameter $\gamma$, triad
resonant interactions are more strongly damped, as is the case for single 
frequency forcing.
However, the situation is more 
complicated here since the modified triad resonant
condition depends on $\gamma$ (actually it depends on the driving amplitude $f$,
which is approximately equal to $\gamma$ at onset).   For example, larger values
of $\gamma$ can make the angle $\theta_{jl}^{(r)}$ at the triad resonance
closer to $\pi/2$.  Figure \ref{fig-gc_100_15} shows the function $g(c_{jl})$
for the same values of parameters as in Fig.~\ref{fig-gc_100_10} except the
value of $\gamma$ is raised to $\gamma=0.15$.

\subsection{Pattern selection near onset}
\label{sec:pattern_selction}

A qualitative analysis of pattern selection for the two frequency case
was already advanced by \cite{re:muller93}, based on general symmetry
considerations and a typical shape of the nonlinear coupling function
$g(c_{jl})$. The derivation of this function given in the previous
section allows us to obtain quantitative predictions for the regions in
parameter space in which patterns of a given symmetry minimize the
appropriate Lyapunov function for this problem. Since
Eq. (\ref{eq:SWAE_scaled}) 
is of gradient form
$1/\gamma \partial_T A_j = -\partial \sF/\partial A_j^*$, a Lyapunov
function $\sF$ can be defined as,
\be
\sF = - \sum_{j=1}^{N} |A_j|^2 + \frac{1}{2}\sum_{j=1}^{N} |A_j|^2
     \left(|A_j|^2 + \sum_{l=1(l\ne j)}^N g(c_{jl}) |A_l|^2\right).
\ee
Since
\be
\frac{d\sF}{dT} = \sum_{j=1}^{N} \left(\frac{\partial \sF}{\partial A_j}
                                       \partial_T A_j
                                      +\frac{\partial \sF}{\partial A_j^*}
                                       \partial_T A_j^* \right)
                = - \frac{2}{\gamma}\sum_{j=1}^{N} |\partial_T A_j|^2 \le 0,
\ee
the only possible limiting cases of such a dissipative system, in the limit
$T \rightarrow \infty$, are stationary states for the amplitudes $A_j$.  Only
the states which correspond to local minima of the Lyapunov function are
linearly stable.

Apart from the trivial solution of $A_j =0$ for $j=1, \cdots, N$,
Eq.~(\ref{eq:SWAE_scaled}) has a family of stationary solutions differing in the
total number of standing waves $N$ for which $A_j \ne 0$.  By considering
the case in which the magnitudes of all standing waves are the same, i.e.
$|A_j| = |A|$, Eq.~(\ref{eq:SWAE_scaled}) has the following solutions,
\be
\label{eq:SWA_solution}
|A_j| = |A| = \left(1+\sum_{l=1(l\ne j)}^{N} g(c_{jl})\right)^{-1/2}.
\ee
The values of the Lyapunov function for these solutions are,
\be
\sF = - \frac{N}{2} |A|^2 = - \frac{N/2}{1+\sum_{l=1(l\ne j)}^{N} g(c_{jl})}.
\ee

We shall only consider pattern structures for which the angle between
any two adjacent wavevectors $\vk_j$ and $\vk_{j+1}$ is the same and amounts
to $\pi/N$ (regular pattern).  We summarize our results concerning
regular patterns in Fig.~\ref{fig-r_phi_2f_selection}.  Note that our
results only apply to the subharmonic response to the driving force,
which corresponds to the side of bicritical line with smaller
amplitude ratio $r$.  Different regions are labeled by the pattern
structure that has the lowest value of the Lyapunov function $\sF_N$
except the region shown in black, in which the self-interaction 
coefficient $g(1)$ is negative.
Fig.~\ref{fig-r_phi_2f_selection}(a) uses all the appropriate fluid
and forcing parameters from M\"uller's experiment (M\"uller 1993). 
As shown in Fig.~\ref{fig-r_phi_2f_selection}(a),
hexagonal or triangular patterns have the lowest value of Lyapunov
function in a region around $\phi=\pi/2$ and close to the bicritical
line while square pattern is stable near the bicritical line and 
around $\phi=3\pi/2$ (or $-\pi/2$).  This result is in qualitative
agreement with M\"uller's experiment although our theory predicts
a smaller stable hexagonal/triangular region for this set of parameter
values.  In a region close to the bicritical line and with 
$0 \le \phi \le \pi/2$, differences in the values of Lyapunov 
function for different patterns become smaller, and most stable
patterns of different symmetries (including quasicrystalline ones)
correspond to neighboring smaller parameter regions.  It is possible
that the preferred pattern may not be selected due to such small differences in 
Lyapunov function for patterns of
different symmetries, or there is no selection at all.  
This result is in partial agreement with M\"uller's experiment,
where he found disordered patterns in a parameter region overlapping
with the above-mentioned region.

Fig.~\ref{fig-r_phi_2f_selection}(b-d) uses the same parameter 
values except the fluid viscosity which is changed from
$\nu = 0.20\>\mbox{cm}^2/\mbox{s}$ to $0.15\>\mbox{cm}^2/\mbox{s}$, 
$0.10\>\mbox{cm}^2/\mbox{s}$,
and $0.05\>\mbox{cm}^2/\mbox{s}$ respectively.  
We note that as $\nu$ decreases (smaller viscous damping), the
hexagonal/triangular region becomes larger and the center of this
region is also shifted towards a larger value of $\phi$.  For
$\nu=0.05\>\mbox{cm}^2/\mbox{s}$, the hexagonal/triangular region reaches 
larger values of $r>0.60$. Currently there are no experimental
results known to us that can be used to test our predictions
as the damping parameter varies.

\section{Discussion and summary}
\label{sec:discussion}

\begin{figure}[t]
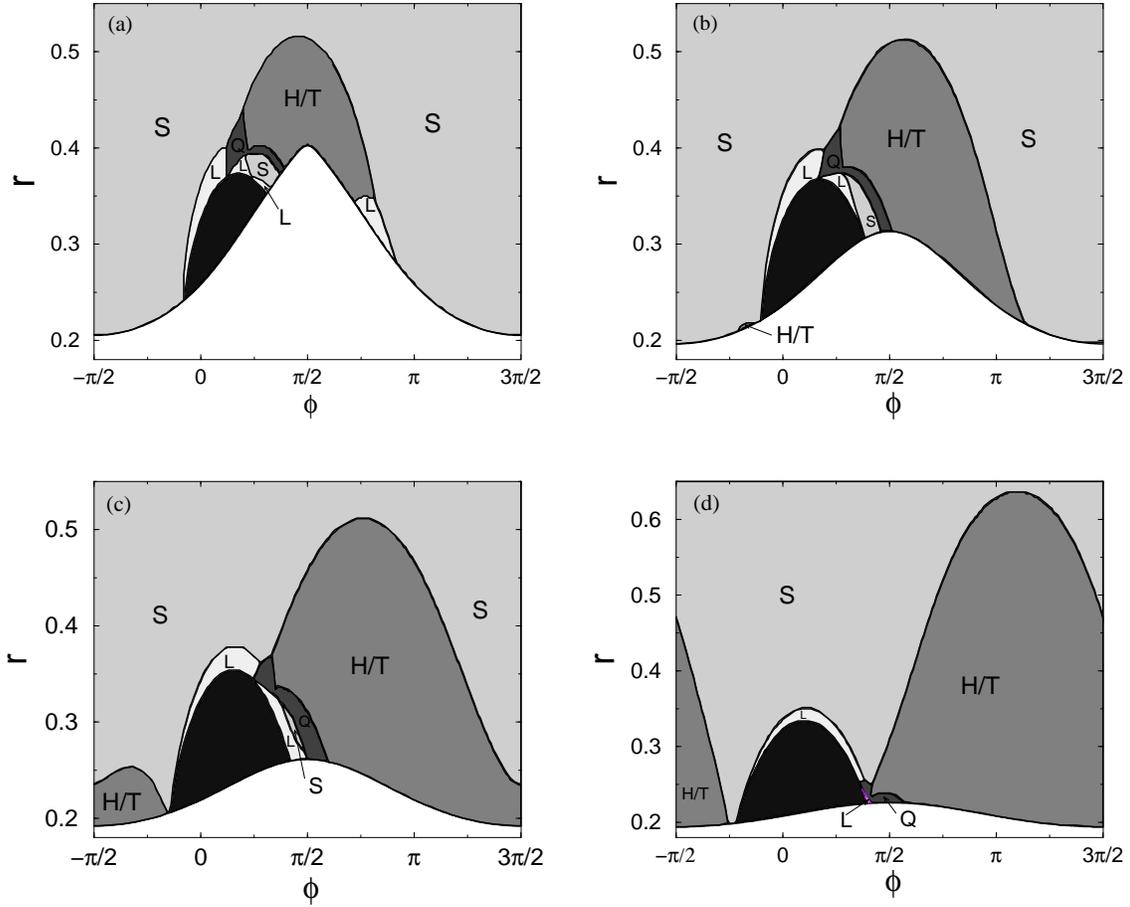

\noindent\hbox{
\psfig{figure=wz6.fig5a,width=3.0in}
\psfig{figure=wz6.fig5b,width=3.0in}
}
\\
\hbox{
\psfig{figure=wz6.fig5c,width=3.0in}
\psfig{figure=wz6.fig5d,width=3.0in}
}
\caption{Regions labeled by the pattern structure that has the lowest value 
of the Lyapunov function: L -- lines, S -- square, H/T -- hexagon/triangle,
and Q -- quasicrystalline patterns (8-fold, 10-fold, 12-fold, ...).  
The black region indicates where the self-interaction coefficient $g(1)$ is 
negative.  
Figure (a) is calculated using parameters of M\"uller's experiment: 
$\rho=0.95\>\mbox{g/cm}^3$, $\Gamma=20.6\>dyn/cm$,
$\nu=0.20\>\mbox{cm}^2/\mbox{s}$, $\omega_0/2\pi=27.9\>\mbox{Hz}$, 
$2\pi/k_{12s}=0.72 \mbox{cm}$, and $2\pi/k_{12h}=17.0\>\mbox{cm}$.   
Figure (b), (c), and (d) are calculated using the same
values of $\rho$, $\Gamma$, $\omega_0$, $k_{12s}$, and $k_{12h}$ as 
(a) but
$\nu=0.15 \mbox{cm}^2/\mbox{s}$,
$\nu=0.1 \mbox{cm}^2/\mbox{s}$, and
$\nu=0.05 \mbox{cm}^2/\mbox{s}$ respectively.  Note the different 
$r$ scale used for (d).}
\label{fig-r_phi_2f_selection}
\end{figure}

We comment further on the approximations considered to obtain
our main results presented in Fig.~\ref{fig-r_phi_2f_selection}.
In the quasi-potential approximation, terms of the order of
$\gamma^{3/2}$ or higher have been neglected. Consistent with
that, we use approximate linear solutions, correct up to order
$f$ or $\gamma$ (the perturbation expansion for the linear solutions)
in deriving the SWAEs (Eq.~(\ref{eq:SWA_two1})).
On the other hand, we included terms of the order of $f^2$, $f\gamma$,
or $\gamma^2$ in the linear stability analysis to obtain the
bicritical line.  We considered these terms because of the
large coefficients for such terms in the perturbation expansion for the
bicritical line in order to compare our results with M\"uller's
experiment which used relatively large values of $\gamma$ (especially
for the case of harmonic response $\gamma_h=0.33$).
A fully consistent calculation correct up to the order of $\gamma^2$
will be much more difficult. Based on the qualitative agreement
of the bicritical line we obtained with experiments, we expect
that a fully consistent calculation would change our results
quantitatively, but not qualitatively.

In summary,
analytical results of bicritical lines $r_b(\phi)$ are obtained by using a 
multiscale perturbation expansion.  The results of $r_b(\phi)$ for frequency
ratio 1/2 agree qualitatively with M\"uller's experimental results.  
We also derive the SWAEs for frequency ratio $1/2$ (for the case of subharmonic
response).  We found that the triad resonance condition is modified due the
presence of the second frequency component.  For certain values of the 
relative phase
difference between the two forcing components, we found that $\theta_{jl}^{(r)}$
becomes close to $90^{\circ}$.  As a result, square patterns become unstable in
this parameter region.   Even though quadratic terms are prohibited
for subharmonic responses, hexagonal or triangular patterns can be stabilized
with the presence of the second frequency component, which is
in agreement with experiments.  We also studied
pattern selection for different values of the damping parameters, and
found that hexagonal/triangular patterns are stabilized in a larger
region for smaller values of the damping parameter.  This is a
prediction that awaits experimental verification.

\begin{acknowledgments} 
This research has been
supported by the U.S.  Department of Energy, contract No.  DE-FG05-95ER14566,
and also in part by the Supercomputer Computations Research Institute, which is
partially funded by the U.S.  Department of Energy, contract No.
DE-FC05-85ER25000.
\end{acknowledgments}

\newpage

\newpage

\newpage

\newpage

\newpage

\end{document}